\documentclass[twocolumn,showpacs,preprintnumbers,amsmath,epsfig,apsrev,aps,prb]{revtex4}

\usepackage{graphicx}
\usepackage{graphics}
\usepackage{dcolumn}
\usepackage{bm}

\begin{document}
\preprint{}

\title{Magnetic excitations of the gapped quantum spin dimer antiferromagnet Sr$_{3}$Cr$_{2}$O$_{8}$}

\author{D. L. Quintero-Castro$^{1,2}$}
\email[]{diana.quintero_castro@helmholtz-berlin.de}
\author{B. Lake$^{1,2}$}
\author{E. M. Wheeler$^{1}$}
\author{A.T.M.N. Islam$^{1}$}
\author{T. Guidi$^{3}$}
\author{K. C. Rule$^{1}$}
\author{Z. Izaola$^{1}$}
\author{M. Russina$^{1}$}
\author{K. Kiefer$^{1}$}
\author{Y. Skourski$^{4}$}
\author{T. Herrmannsd\"orfer$^{4}$}

\affiliation{$^{1}$Helmholtz Zentrum Berlin f\"ur Materialien und Energie, D-14109 Berlin, Germany \\
$^{2}$Institut f\"ur Festk\"orperphysik, Technische Universit\"at Berlin, D-10623 Berlin, Germany\\
$^{3}$ISIS Facility, Rutherford Appleton Laboratory, Chilton, Didcot, Oxon OX11 0QX, United Kingdom\\
$^{4}$Hochfeld Magnetlab, Forschungszentrum Dresden-Rossendorf EV, D-01314 Dresden, Germany}

\date{\today}

\begin{abstract}
Sr$_{3}$Cr$_{2}$O$_{8}$ consist of a lattice of spin-$1/2$ Cr$^{5+}$ ions, which form hexagonal bilayers and which are paired into dimers by the dominant antiferromagnetic intrabilayer coupling. The dimers are coupled three-dimensionally by frustrated interdimer interactions. A structural distortion from hexagonal to monoclinic leads to orbital order and lifts the frustration giving rise to spatially anisotropic exchange interactions. We have grown large single crystals of Sr$_{3}$Cr$_{2}$O$_{8}$ and have performed DC susceptibility, high field magnetisation and inelastic neutron scattering measurements. The neutron scattering experiments reveal three gapped and dispersive singlet to triplet modes arising from the three twinned domains that form below the transition thus confirming the picture of orbital ordering. The exchange interactions are extracted by comparing the data to a Random Phase Approximation model and the dimer coupling is found to be $J_{0}=5.55$\,meV, while the ratio of interdimer to intradimer exchange constants is $J'/J_{0}=0.64$. The results are compared to those for other gapped magnets.
\end{abstract}

\maketitle

\section{Introduction}
Over the last decade, many exotic quantum phenomena have been observed in spin dimer systems. For example TlCuCl$_{3}$, which consists of spin-1/2 dimers coupled in a three-dimensional lattice, has a spin-singlet ground state and a gapped dispersive one-magnon excited state. Application of a magnetic field causes Bose-Einstein condensation of magnons at a critical field $H_{c}$, while for higher fields magnetization increases smoothly to saturation \cite{Rueggnature, NikuniPRL84}. BaCuSi$_{2}$O$_ {6}$ consists of Cu$^{2+}$ ions arranged on a bilayer of square lattices with unfrustrated antiferromagnetic couplings. The intrabilayer coupling is dominant so that the system can be regarded as a square lattice of dimers. The ground state is again a singlet, the excitations are dispersive gapped magnons \cite{SasagoPRB55, RueggPRL98} and as for TlCuCl$_{3}$, magnetization increases smoothly above a critical field. The spin-$1/2$ Shastry-Sutherland antiferromagnet SrCu$_{2}$(BO$_{3}$)$_{2}$ consists of spin-$1/2$ dimers arranged at right angles to each other so that the interdimer couplings are competing. The ground state is again a singlet and the excitations form a gapped mode whose dispersion bandwidth is suppressed by frustration \cite{RueggPRL93}. Magnetization is characterized by a critical field and for higher fields a series of magnetization plateaux are found where the condensed magnons form superlattices with spontaneous breaking of translation symmetry known as Wigner crystallization \cite{KageyamaPRL82}.

Recently a new set of dimerized antiferromagnets with formula A$_{3}$M$_{2}$O$_{8}$ (A=Ba, Sr and M=Mn, Cr) have been investigated. In these compounds the magnetic ion (Mn or Cr) is in the $5^{+}$ oxidation state leading to spin-$1$ in the case of Mn$^{5+}$ and spin-$1/2$ in the case of Cr$^{5+}$. These compounds crystallize in a hexagonal structure with space group $R\bar{3}m$ at room temperature. The magnetic ions lie on hexagonal bilayers where the dominant intrabilayer interaction is antiferromagnetic and gives rise to the dimerization. These dimers are coupled three-dimensionally, both within the bilayers and between bilayers to give a highly frustrated arrangement. Ba$_{3}$Mn$_{2}$O$_{8}$ maintains the hexagonal structure down to the lowest temperatures. Stone $et\, al.$ \cite{StonePRL100} found that the magnetic excitations are gapped, dispersive magnons modes typical of a dimerized magnet. Magnetic exchange constants have been extracted by fitting the single mode approximation to the data \cite{StonePRL100}, and by spin dimer analysis based on a extended H\"{u}ckel tight binding calculation \cite{Koo}. The magnetization shows two magnetic plateaux. The first is due to a triplet condensation into the ground state and the second is a quintet condensation. Critical properties of Ba$_{3}$Mn$_{2}$O$_{8}$ have also been extensively studied by Suh and Samulon \cite{Suharxiv} who find that the critical points can be defined as a Bose Einstein condensation of magnons tuned by an external magnetic field.

In the case of Ba$_{3}$Cr$_{2}$O$_{8}$ and Sr$_{3}$Cr$_{2}$O$_{8}$ the unusual 5$^{+}$ valence state of Chromium results in a single electron in the $3d$ shell. The Cr ion is surrounded by an oxygen tetrahedron and this crystal field splits the $3d$ levels so that electron occupies the doubly degenerate lower lying $e$ orbitals. The Cr$^{5+}$ ion in this environment is thus Jahn-Teller active and a Jahn-Teller distortion lifts both the orbital degeneracy and magnetic frustration. This distortion has been confirmed in Ba$_{3}$Cr$_{2}$O$_{8}$ by neutron powder diffraction where additional peaks corresponding to the monoclinic $C2/c$ symmetry were found \cite{KofuPRLJan}. In addition three gapped and dispersive excitation branches were observed in inelastic neutron scattering, corresponding to the three twinned domains that form below the transition \cite{KofuPRLJan}. The magnetic excitations were fitted by a random phase approximation model (RPA) and the interdimer couplings were shown to be spatially anisotropic as would be expected in the monoclinic phase \cite{KofuPRLJan}. The magnetization was also consistent with gapped excitations showing zero magnetization up to $H_{c}=12.5$\,T and then a smooth increase up to saturation at $H_{s}=23.6$\,T \cite{KofuPRLMay, AczelPRB2009}. A combination of neutron diffraction, heat capacity, magnetocaloric effect and magnetic torque measurements were performed at high magnetic fields to map out the phase diagram as a function of field and temperature around the critical fields \cite{KofuPRLMay, AczelPRB2009}. The extracted critical exponents were found to be consistent with theoretical predictions for a three-dimensional Bose-Einstein condensation at $H_{c}$\, although not at $H_{s}$.

Here we investigate the magnetic properties of Sr$_{3}$Cr$_{2}$O$_{8}$, which has very recently also been found to show Bose-Einstein condensation behaviour \cite{AczelLANL}. 
The room temperature crystal structure of Sr$_{3}$Cr$_{2}$O$_{8}$ was solved in $1989$ by Cuno and M\"{u}ller and found to be hexagonal with space group $R\bar{3}m$ \cite{Cuno1989}. A decade later the thermodynamic properties were investigated by Jacob \cite{Jacob2000}, but their samples were shown by XANES to have a mixed valence state. Much more recently Singh $et\, al.$ \cite{Singh} measured DC magnetic susceptibility and heat capacity on polycrystalline samples. The susceptibility revealed a spin singlet ground state and could be fitted by a model of weakly coupled dimers with intradimer exchange constant of 5.34\,meV at base temperature, while the specific heat confirmed the absence of long range magnetic order in this compound. Chapon $et\, al.$ performed powder neutron diffraction and found that Sr$_{3}$Cr$_{2}$O$_{8}$ undergoes a structural phase transition below 275\,K to a monoclinic structure with space group $C2/c$ \cite{Chapon}. The transition is thought to be a cooperative Jahn-Teller distortion that lifts the degeneracy of the $e$ orbitals, stabilizing the $3z^{2}-r^{2}$ orbital at lower energy than the $x^{2}-y^{2}$ and giving rise to antiferro orbital ordering. This ordering alters the magnetic exchange paths and lifts the magnetic frustration. At room temperature in the hexagonal phase the lattice parameters are $a_{h}=b_{h}=5.57$\,\AA, $c_{h}=20.17$\,\AA\ while at $1.6$\,K in the monoclinic phase the lattice parameters are: $a_{m}=9.66$\,\AA, $b_{m}=5.5437$\,\AA, $c_{m}=13.7882$\,\AA, $\beta=103.66$$^\circ$. The subscript `$h$' denotes hexagonal notation and `$m$' monoclinic.

\begin{figure}[htb!]
        \includegraphics[width=0.5\textwidth]{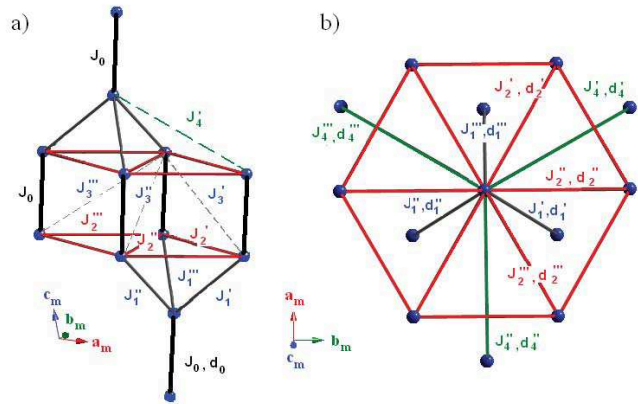}
        \caption{(Color online) The low temperature (monoclinic) crystal structure of Sr$_{3}$Cr$_{2}$O$_{8}$ giving the magnetic Cr$^{5+}$ ions only. (a) Shows the bilayer structure, (b) Shows the `hexagonal' plane. The exchange interactions and Cr$^{5+}$-Cr$^{5+}$ distances are labeled on the diagram.}
    \label{fig:structure}
\end{figure}

Figure \ref{fig:structure} shows the network of Cr$^{5+}$ ions forming the structure of Sr$_{3}$Cr$_{2}$O$_{8}$. The hexagonal layers lie in the $(a_{h},b_{h})$-plane and are paired into bilayers in the $c_{h}$ direction. These bilayers are in turn stacked along $c_{h}$ in a repeated $ABCABC$ pattern. The shortest Cr$^{5+}$-Cr$^{5+}$ distance is the first neighbor intrabilayer distance (labeled $d_{0}$ in figure \ref{fig:structure}a) and the corresponding coupling constant is $J_{0}$. In the high temperature hexagonal phase the Cr$^{5+}$ ions, within the hexagonal layers, are separated by distance $d_{2}$ and are coupled by the exchange constant $J_{2}$. However below the structural distortion the hexagonal distances become unequal (labeled $d_{2}'$, $d_{2}''$, $d_{2}'''$ in Fig \ref{fig:structure}) corresponding to three different exchange interactions $J_{2}'$, $J_{2}''$, and $J_{2}'''$.  The same is true for the third neighbor intrabilayer exchange constant and the interbilayer interactions. These spatially anisotropic exchange interactions would result in crystal twinning since there are three possible ways in which the hexagonal symmetry can be broken to form the monoclinc structure.


In this paper we investigate the magnetic properties of Sr$_{3}$Cr$_{2}$O$_{8}$. We study the system using DC susceptibility, high field magnetization and inelastic neutron scattering on both powder and single crystal samples. We show that Sr$_{3}$Cr$_{2}$O$_{8}$ is indeed a dimerized magnet with $J_{0}=5.55$\,meV and substantial three-dimensional interdimer couplings. The neutron scattering data reveal three excitation modes corresponding to the three twinned domains thus confirming the Jahn-Teller distortion and the spatially anisotropic exchange interactions that arise at this transition. The data are compared to a RPA model and the exchange constants are determined. The results are compared to those for Ba$_{3}$Mn$_{2}$O$_{8}$ and Ba$_{3}$Cr$_{2}$O$_{8}$ and are discussed in relation to other gapped magnets.

\section{Experimental details}

Single crystalline samples of Sr$_{3}$Cr$_{2}$O$_{8}$ were growth in the Crystal Laboratory at the Helmholtz Zentrum Berlin f\"{u}r Materialien und Energie (HZB) in Berlin, Germany. The initial powders were synthesized by a solid state reaction. Stoichiometric amounts of SrCO$_{3}$ ($99.994\%$ purity, Alfa Aesar) and Cr$_{2}$O$_{3}$ ($99.97\%$ purity, Alfa Aesar) were mixed thoroughly, pressed into a pellet and heated first for $24$ hours at $850$$^\circ\text{C}$ and then for $24$ hours at $1250$$^\circ\text{C}$ (with intermediate grindings). The powders were quenched to avoid the formation of other phases like SrCrO$_{4}$ and Sr$_{10}$Cr$_{6}$O$_{24}$(OH)$_{2}$ and were then pressed into a feed rod using a cold isostatic press (EPSI). The single crystal growth was performed by the floating zone technique, using a high temperature optical floating zone furnace (Crystal Systems Inc. Model FZ-T-10000-H-VI-VPO). An atmosphere of synthetic air was used to maintain the correct level of oxygen needed for the high oxidation state of the Cr$^{5+}$ and to reduce the moisture level to prevent the growth of other phases. The resulting samples were typically cylindrical with diameter $\approx6$\,mm and length $35-50$\,mm. Further details of the growth of Sr$_{3}$Cr$_{2}$O$_{8}$ will be published elsewhere \cite{Nazmul}.

	The static magnetic susceptibility and low field magnetization of Sr$_{3}$Cr$_{2}$O$_{8}$ were measured using a superconducting quantum interference device (MPMS, Quantum Design) at the Laboratory for Magnetic Measurements, HZB. The samples had a volume of $3$\,mm$^{3}$ and the data were collected with a magnetic field applied both parallel and perpendicular to $c_{h}$. The susceptibility measurements were performed using a field of $3$\,T over a temperature range of $2-300$\,K. The high-field magnetization of Sr$_{3}$Cr$_{2}$O$_{8}$ was measured using a pulsed inductive magnetometer at the Hoch-Field-Institute in Dresden. The measurements took place over a field range of $0-65$\,T and a variety of temperatures from $1.5$ to $50$\,K, again both field directions ($H\bot c_{h}$ and $H // c_{h}$) were measured.	

	The magnetic excitation spectrum was investigated using inelastic neutron scattering. For an overview of the excitation spectrum, powder measurements were performed on the cold multidisc-chopper neutron time-of-flight spectrometer $V3-NEAT$ at the BER II reactor, HZB. The sample (mass $=9.5$\,g) was inserted into a cylindrical aluminum container and mounted inside a standard orange cryostat. Data were collected for incident wavelengths (energies) of $2$\,\AA ($20.45$\,meV), $2.5$\,\AA ($13.09$\,meV) and $3.2$\,\AA ($7.99$\,meV), resulting in a resolution at the elastic line (at $5.3$\,meV energy transfer) of $1.437$\,meV ($1.053$\,meV), $0.742$\,meV ($0.448$\,meV), $0.442$\,meV ($0.187$\,meV) respectively. All the data were collected at $2$\,K and each scan was counted for approximately $20$ hours. A vanadium standard was used for the detector normalization and empty can measurements were used for the background subtraction.
	
	For a detailed study of the magnetic dispersion, two types of single crystal inelastic neutron scattering measurements were performed. Firstly, using MERLIN \cite{Merlin}, a direct geometry time-of-flight spectrometer at the ISIS facility, Rutherford Appleton Laboratory, UK and secondly the cold-neutron triple-axis-spectrometer V2-FLEX at HZB. The MERLIN experiment allowed the magnetic excitations to be measured over a large range of energy and wavevector transfer and provided an overview of the spectrum. Three single crystals of total mass $10$\,g, were co-aligned with the $c_{h}$ axis vertical and cooled in a closed cycle cryostat to a base temperature of 6 K. A Fermi chopper was phased to select neutrons with an incident energy of $10$\,meV and was spun at a frequency of $100$\,Hz to achieve a resolution at the elastic line of $0.37$\,meV. An omega scan was performed where the sample was rotated about the vertical axis in steps of $1$$^\circ$ covering an angular range of $70$$^\circ$ with a counting time of $2$ hours per step. A vanadium standard was used to normalize the individual measurements for detector efficiency and then the full data set was combined using the program Horace \cite{Horace} to give the excitation spectrum over a large volume of reciprocal space and energy transfer.
	
	The measurements performed on V2-FLEX were used to obtain high resolution data for accurate mapping of the dispersion relations. Two separate experiments were performed. For the first, a single crystal (mass\,$=3.7$\,g) was oriented with the ($h_{h},0_{h},\ell_{h}$) plane as the instrumental scattering plane. For the second experiment two co-aligned single crystals (total mass$=7.7$\,g) were used to investigate the ($h_{h},h_{h},\ell_{h}$) plane. An orange cryostat was used to cool the sample and most of the measurements were performed at $2$\,K. A vertically-focusing Pyrolitic graphite (PG) monochromator and horizontally-focusing PG analyzer were used to select incident and final energies. A nitrogen-cooled beryllium filter was used between the sample and analyzer to remove  $\lambda/2$ and higher order reflections from the monochromator. The instrument was operated in the W-configuration $(+ - +)$ and the collimator settings were guide-$60'$-open-open. The measurements were performed with fixed final wavevectors of $k_{f}=1.2$\,\AA$^{-1}$ (2.98\,meV) and $k_{f}=1.55$\,\AA$^{-1}$ (4.98\,meV).  The resolution of the instrument, determined from the full-width-half-maximum (FWHM) of the incoherent scattering was $0.09$\,meV for $k_{f}=1.2$\,\AA$^{-1}$ and $0.12$\,meV for a $k_{f}=1.55$\,\AA$^{-1}$. Constant-wavevector scans were used to map out the magnon dispersions from $0-8$\,meV in steps of $0.05$\,meV. The average counting time was 4 minutes per point.

\section{Results and Discussion}

\subsection{Static Susceptibility}

\begin{figure}[htb!]
\centering
		\includegraphics[width=0.5\textwidth]{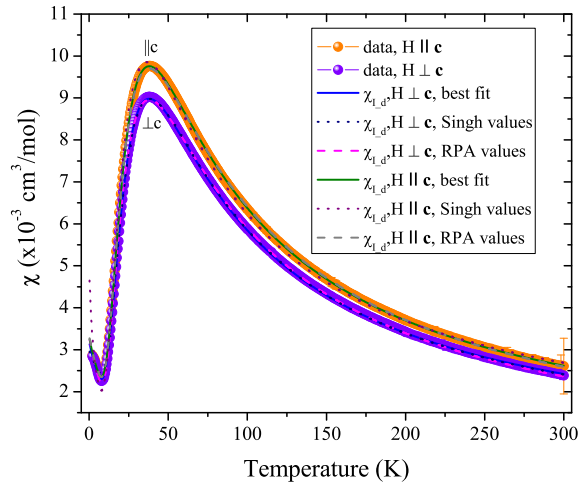}
		\caption{(Color online) Static susceptibility data measured using the SQUID magnetometer for field applied both parallel and perpendicular to $c_{h}$ displayed along with the fitted interacting dimer model given by Eq. (1).}
\label{fig:susceptibility}
\end{figure}

The susceptibility of Sr$_{3}$Cr$_{2}$O$_{8}$ is displayed in figure \ref{fig:susceptibility}. The figure shows the results of the susceptibility with the field applied both perpendicular and parallel to $c_{h}$, the two curves are slightly different suggesting a weakly anisotropic $g$ tensor as found for Ba$_{3}$Cr$_{2}$O$_{8}$ \cite{AczelPRB2009}. The results reveal a broad maximum at $38$\,K and a sharp drop toward zero at lower temperatures as found in the previous powder measurement \cite{Singh}. Such a susceptibility curve is characteristic of a dimerized magnet with a singlet ground state and gapped magnetic excitations. The susceptibility rises again at very low temperatures presumably due to paramagnetic impurities. Note however that this paramagnetic tail is smaller than in the previous measurements \cite{Singh} suggesting a higher quality sample. Altogether we expect three contributions to the susceptibility,
\begin{equation}
	\chi_{obs}=\chi_{0}+\chi_{imp}+\chi_{interacting-dimer}.
\end{equation}
The temperature independent term, $\chi_{0}$, is due to Van Vleck paramagnetism and diamagnetic core susceptibility and is expected to give a small contribution $\sim 10^{-4}$\,cm$^{3}/$mol. The second term, $\chi_{imp}$, is due to paramagnetic impurities which are responsible for the Curie tail at low temperatures and has the form:
\begin{equation}
	\chi_{imp}=\frac{C_{imp}}{T-\theta _{imp}},
\end{equation}
where $C_{imp}$ is the Curie constant of the impurities and $\theta_{imp}$ is their Curie-Weiss temperature. The third term, $\chi_{interacting-dimer}$, is due to the coupled dimer arrangement proposed for Sr$_{3}$Cr$_{2}$O$_{8}$ and is modeled by
\begin{equation}
	\chi_{interacting-dimer}=\frac{3C/T}{(3+e^{(J_{0}/k_{B}T)}+J'/T)}.
\end{equation}
This expression is described in references \onlinecite{Singh} and \onlinecite{Johnston1997}. It is derived from the Bleaney-Bowers equation for spin-$1/2$ dimers coupled by Heisenberg intradimer exchange constant $J_{0}$, and $J'$ is the effective field on each spin due to its couplings with the neighboring spins. In the case of unfrustrated interdimer couplings $J'$ is simply the sum of the modulus of these couplings, which for Sr$_{3}$Cr$_{2}$O$_{8}$ would be $J'=(|J_{1}'|+|J_{1}''|+|J_{1}'''|)+2(|J_{2}'|+|J_{2}''|+|J_{2}'''|)+2(|J_{3}'|+|J_{3}''|+|J_{3}'''|)+(|J_{4}'|+|J_{4}''|+|J_{4}'''|)$. However if the interactions are competing due to frustration they act against each other reducing the effective size of $J'$. The constant $C$ is the Curie constant of this system.
	
The measured susceptibility data was fitted to Eqn.$(1)$ and good agreement was found over the full temperature range, see Fig. \ref{fig:susceptibility}. The fitted parameters for the data with the field perpendicular to the $c_{h}$ axis are  $\chi_{0}=2.32(5)\times10^{-4}$\,cm$^{3}/$mol, $C_{imp}=0.0465(5)$\,cm$^{3}$K$/$mol,  $\theta_{imp}=-15.7(2)$\,K, $C=0.875(2)$\,cm$^{3}$K$/$mol, $J_{0}=5.517(2)$\,meV, $J'=1.22(5)$\,meV. The value of $C$ corresponds to an effective moment of $1.8(1) \mu_{B}/$Cr$^{5+}$, which is close to the value of $1.72\mu_{B}$ expected for spin-$1/2$ Cr$^{5+}$ ions assuming a $g$-factor of $1.98$ taken from a previous EPR measurement \cite{Gaft2000}. The values obtained from fitting the susceptibility data with the field parallel to the $c_{h}$ axis are: $\chi_{0}=1.76(5)\times10^{-4}$\,cm$^3/$mol, $C_{imp}=0.0481(6)$\,cm$^3$K$/$mol,  $\theta_{imp}=-15.3(3)$\,K, $C=0.987(2)$\,cm$^3$K$/$mol, $J_{0}=5.512(3)$\,meV, $J'=1.97(6)$\,meV, giving $\mu_{eff}= 1.9(1)\mu_{B}/$Cr$^{5+}$.

Our values for the intradimer and interdimer exchange interactions can be compared to the values of $J_{0}=5.17(1)$\,meV and $J'= 0.5(2)$\,meV found previously by Singh $et\, al.$ \cite{Singh}. While our fitted intradimer exchange constant is in good agreement, our value of the interdimer coupling is much greater. If the values of $J_{0}$ and $J'$ are however fixed to those of Singh $et\, al.$ \cite{Singh} we obtain a fit that is almost as good as shown in Fig. \ref{fig:susceptibility}. Furthermore the values of $J_{0} = 5.5512$\,meV and $J'=3.583$\,meV, obtained by fitting our inelastic neutron scattering data to a RPA model as described later in the text, where the size of $J'$ was found to be much larger, gave equally good agreement to the data. Altogether our susceptibility agrees well with a model of interacting spin-$1/2$ dimers and gives an intradimer exchange constant of $J_{0}\approx5.5$\,meV. However, it is unable to give accurate information about the interdimer exchange coupling and for that information we turn firstly to magnetization and then inelastic neutron scattering.

\subsection{High field magnetization}

\begin{figure}[htb!]
	\centering
		\includegraphics[width=0.5\textwidth]{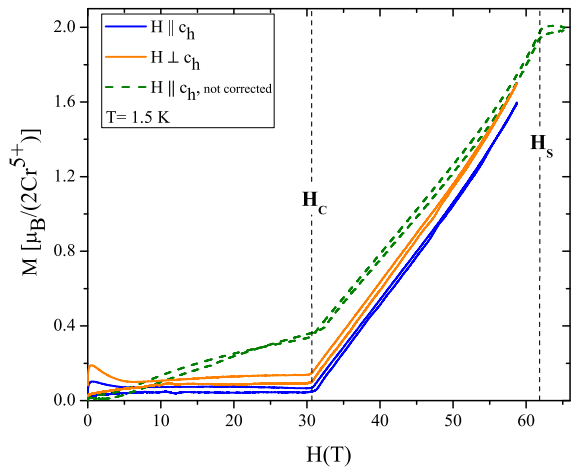}
		\caption{(Color online) Magnetization per 2Cr$^{5+}$ ions as a function of applied magnetic field. This measurement was performed using a pulsed magnet at base temperature $T=1.5$\,K and field applied both parallel and perpendicular to $c_{h}$. The data represented by the green line is not background corrections.}
	\label{fig:magnetization}
\end{figure}

The high field magnetization of Sr$_{3}$Cr$_{2}$O$_{8}$ reveals three distinct regions, see Fig. \ref{fig:magnetization}. Below the critical field of $H_{c}=30.9(4)$\,T there is zero magnetization (the gradual increase in magnetization observed in the 65 T measurement (green line) is because this data has not been background corrected). Between $H_{c}$ and $H_{s}=61.9(3)$\,T the magnetization increases rapidly and smoothly, while above $H_{s}$ a magnetization plateau is observed at $M_{s}=1.97(3)\mu_{B}/2$Cr$^{5+}$. This is identified as the saturation plateau since it is very close to the theoretical value of $2g_{s}\mu_{B}S=1.98\mu_{B}/2$Cr$^{5+}$ (assuming $g=1.984$).	
	
	The observed magnetization can be explained within the proposed interacting dimer model of Sr$_{3}$Cr$_{2}$O$_{8}$. In this model the ground state is a singlet resulting in no magnetization at low magnetic fields. The excitations would form a gapped magnon mode which disperses between an upper and lower energy. The magnons possess a spin of $S=1$ and their triplet degeneracy is lifted by an applied magnetic field. The critical field corresponds to a Zeeman energy such that the gap is closed by the $S_{z}=1$ magnons which start to populate the ground state; from this we can deduce that $E_{gap}=g_{s}\mu_{B}H_{c}=3.54(5)$\,meV.  The saturation field corresponds to the ground state being fully occupied by excited dimers and occurs when the Zeeman energy is equal to the upper edge of the dispersion, thus giving $E_{upper}=g_{s}\mu_{B}H_{s}=7.09(4)$\,meV.
	
	The magnetization data are therefore consistent with the model of interacting spin-$1/2$ dimers and suggests that the excitations are centered around $5.3$\,meV and disperse over a bandwidth of $3.55$\,meV. Since in unfrustrated dimerized magnets the bandwidth is equal to $J'$ (the sum of the modulus of the interdimer exchange constants) the significant bandwidth revealed by the magnetization suggests substantial interdimer couplings with $J'\approx3.55$\,meV. Finally it should be noted that the absence of any intermediate magnetization plateau is in contrast to the findings for some other gapped magnets that are highly frustrated e.g. SrCu$_{2}$(BO$_{3}$)$_{2}$ which has steps in its magnetization. This result is however in agreement with related compound Ba$_{3}$Cr$_{2}$O$_{8}$.

\subsection{Powder inelastic neutron scattering}

\begin{figure}[htb!]
	\centering
		\includegraphics[width=0.5\textwidth]{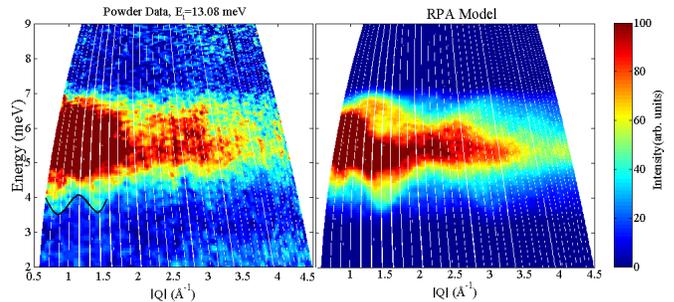}
		\caption {(Color online) The left panel shows the inelastic neutron scattering data measured on a powder sample. The data was collected on NEAT at $2$\,K and the incident energy was $13.08$\,meV. Black line is a guide to the eye to highlight the lower edge of the excitation band. The right panel shows the RPA simulation over the same energy and wavevector ranges.}
	\label{fig:powder}
\end{figure}

\begin{figure}[htb!]
	\centering
		\includegraphics[width=0.5\textwidth]{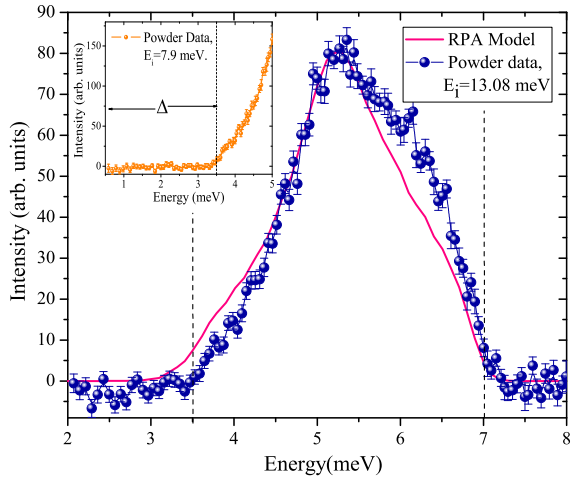}
		\caption{(Color online) The data (blue circles) and RPA simulation (magenta line) from Fig. \ref{fig:powder}, integrated over the wavevector range $1.5<\left|Q\right|<3.5$\,\AA$^{-1}$ \ and plotted as a function of energy. The dashed lines indicate the bandwidth of the excitations. The inset shows the lower energy region and reveal the clean energy gap ($\Delta$), this data was taken with an incident energy of 7.9\,meV and the cut is over the wavevector range of $0.6<\left|Q\right|<1.5$\,\AA$^{-1}$.}
	\label{fig:bandwidth}
\end{figure}

In order to learn more about the magnetic excitations of Sr$_{3}$Cr$_{2}$O$_{8}$ we performed inelastic neutron scattering measurements on a powder sample. Figure \ref{fig:powder} shows data collected on NEAT at $2$\,K plotted as a function of wavevector transfer $\left|Q\right|$ and energy transfer $E$. The spectrum reveals a band of gapped excitations centered at $\sim$5.5\,meV. The measured neutron scattering intensity reduces with increasing $\left|Q\right|$ confirming the magnetic nature of the excitations. Figure \ref{fig:bandwidth} gives the spectrum integrated over the wavevector range $1.5<\left|Q\right|<3.5$\,\AA$^{-1}$ and plotted as a function of energy, and clearly shows that within resolution, $E_{gap}=3.4(3)$\,meV and $E_{upper}=7.10(5)$\,meV. These results are in good agreement with the values obtained from our magnetization measurements and confirm the picture of interacting dimers giving rise to gapped and dispersive magnon excitations. Additional measurements using a smaller incident energy of $7.99$\,meV were performed to probe the lower energy region more carefully and confirm the complete absence of magnetic scattering below $3.5$\,meV as shown in the inset of Fig.\ref{fig:bandwidth}.

The $\left|Q\right|$ averaging due to the powder sample prevents exact determination of the dispersion relations; however some information can be gained by looking at the lower edge of the excitation band. A minimum in the excitation spectrum occurs at $|Q|=0.87$\,\AA$^{-1}$, which corresponds to the reciprocal lattice point $(1/2_{h},0_{h},1_{h})$ (see black line in Fig.\ref{fig:powder}), and further minima are observed e.g. at $|Q|=1.48$\,\AA$^{-1}$ suggesting that the minima of the dispersion occur at $(h_{h}/2,0_{h},\ell_{h})$ where $h_{h}$ and $\ell_{h}$ are integer, as well as symmetry equivalent positions.

Information about the relative importance of the various magnetic couplings in Sr$_{3}$Cr$_{2}$O$_{8}$ can be extracted from the data by using the first moment sum rule \cite{Hohenberg1974}. For an isotropic spin system where the magnetic ions are coupled by Heisenberg exchange interactions, the first moment $\left\langle E \right\rangle$ of the scattering function, $S(Q,\omega)$, is given by
\begin{align}
	\left\langle E \right\rangle & =\hbar^2\int^{\infty}_{\omega=-\infty} \omega S(\mathbf{Q},\omega)d\omega \nonumber \\
	& \propto-\sum_{i}J_{i}\left\langle \mathbf{S}_{0}\cdot \mathbf{S}_{d_{i}}\right\rangle (1-cos(\mathbf{Q}\cdot \mathbf{d}_{i})).
\end{align}
In this equation $J_{i}$ is the exchange constant coupling the $i^{th}$ nearest neighbor spins, $d_{i}$ is their separation and $\left|\left\langle \textbf{S}_{0}.\textbf{S}_{d_{i}}\right\rangle\right|$ is the two spin correlation function for this pair. From this expression it is clear that each coupling constant $J_{i}$ produces a modulation in the first moment with a periodicity depending on the separation $d_{i}$ of the spins. Thus by fitting this expression to the first moment of the data it should be possible to identify the dominant exchange paths.

\begin{figure}[htb!]
	\centering
		\includegraphics[width=0.5\textwidth]{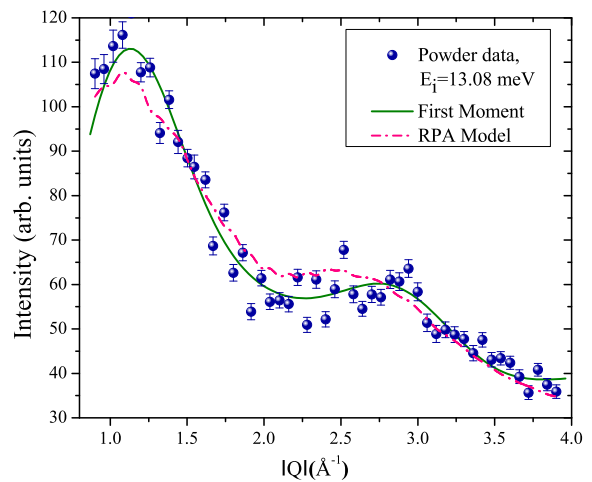}
		\caption {(Color online) The data (blue circles) and RPA simulation (magenta line) from Fig. \ref{fig:powder}, multiplied by energy transfer and integrated over energy. The green line gives the fit to the first moment sum rule, Eqn (6).}
	\label{fig:firstmoment}
\end{figure}

For comparison to the powder measurement of Sr$_{3}$Cr$_{2}$O$_{8}$ it is necessary to take the powder average and include the magnetic form factor of Cr$^{5+}$ , $f_{\mathrm{Cr}^{5+}}(\left|Q\right|)$, such that
\begin{equation}
	\left\langle E \right \rangle \propto \sum_{i} J_{i} \left\langle \mathbf{S}_{0}\cdot \mathbf{S}_{i}\right\rangle \left|f_{\mathrm{Cr}^{5+}}(\left|Q\right|)\right|^{2} \left(1-\frac{sin(\left|Q\right| \left|d_{i}\right| )} {\left|Q\right|\left|d_{i}\right|}\right).
\end{equation}
The possible interactions for the monoclinic phase of Sr$_{3}$Cr$_{2}$O$_{8}$ are illustrated in figure \ref{fig:structure}, however since the set of distances $d_{i}'$, $d_{i}''$ and $d_{i}'''$ ($i\neq0$) are almost equal it is not possible to distinguish between them in the fitting and we therefore replace these distances by the average distance $d_{i}$  ($d_{0}= 3.8448$\,\AA, $d_{1}= 4.3172$\,\AA,  $d_{2}=5.5718$\,\AA, $d_{3}= 6.7696$\,\AA). The expression then becomes
\begin{align}
\left\langle E \right \rangle \propto
-\sum_{i=0}^{3}A_{i}\left|f_{\mathrm{Cr}^{5+}}(\left|Q\right|)\right|^{2}\left(1-\frac{sin(\left|Q\right|\left|d_{i}\right|)}{\left|Q\right|\left|d_{i}\right|}\right).
\end{align}
Where the terms $A_{0}$ and $A_{i}$ are
\begin{align}
A_{0}&=J_{0}\left\langle \mathbf{S}_{0}\cdot \mathbf{S}_{d_{0}}\right\rangle \\
A_{i}&=J_{i}'\left\langle \mathbf{S}_{0}\cdot \mathbf{S}_{d_{i}'}\right\rangle+ J_{i}''\left\langle \mathbf{S}_{0}\cdot \mathbf{S}_{d_{i}''}\right\rangle+J_{i}'''\left\langle \mathbf{S}_{0}\cdot \mathbf{S}_{d_{i}'''}\right\rangle
\end{align}
An expression for the form factor of Cr$^{5+}$ could not be found in the literature, we therefore estimated it by extrapolation from the known form factors for Cr$^{4+}$ and V$^{4+}$ to be
\begin{equation}
	\left|f_{\mathrm{Cr}^{5+}}(\left|Q\right|)\right|^2=\left|Ae^{-as^2}+Be^{-bs^2}+Ce^{-cs^2}+D\right|^{2}
\end{equation}
where $s=\left|Q\right|/4\pi$  and the constants are $A = -0.2602$, $a = 0.03958$, $B = 0.33655$, $b = 15.24915$, $C = 0.90596$, $c = 3.2568$ and $D = 0.0159$.
The data, multiplied by energy transfer and integrated over the energy to give the first moment is shown in Fig. \ref{fig:firstmoment} as a function of wavevector. The theoretical expression in equation $(6)$ was then fitted by varying the quantities $A_{0}$ and $A_{i}$ and the best fit, shown by the green line, corresponds to fitted values $A_{0} = 402(3)$, $A_{1}= -24(5)$, $A_{2}=-24(4)$, $A_{3}=-45(4)$.

The results reveal that, as expected, the dominant exchange constant of Sr$_{3}$Cr$_{2}$O$_{8}$ which is responsible for coupling the spins into dimers is the intrabilayer interaction, $J_{0}$. Indeed the quantity $A_{0}$ is more that an order of magnitude greater than the other $A_{i}$ values, clearly identifying it as the main interaction. Unfortunately it is not possible to obtain the absolute values of the exchange constants using this method as the fitted quantities also depend on the spin correlation functions. Nevertheless it is clear that there are significant interdimer interactions and that Sr$_{3}$Cr$_{2}$O$_{8}$ can be viewed as a network of dimers coupled both along the $c_{h}$ axis and within the distorted `hexagonal' plane. These results can be compared to the inelastic neutron scattering measurements of Chapon $et\, al.$ \cite{Chapon} which were also performed on a powder sample. A gapped band of excitations was observed centered at an energy of $5.46$\,meV in agreement with our data. The FWHM of this band was found to be $2.1$\,meV although values of the gap and upper boundary were not found due to the poorer energy resolution of their measurement.
	
In order to gain more information about the exchange paths in Sr$_{3}$Cr$_{2}$O$_{8}$ it is necessary to measure the magnon dispersions along specific directions in reciprocal space, since only in this way can the set of interactions $J_{i}'$, $J_{i}''$ and $J_{i}'''$ be distinguished from each other and their values accurately determined. This can be achieved by inelastic neutron scattering measurements from a single crystal sample and is the subject of the next section.

\subsection{Single crystal inelastic neutron scattering}

\subsubsection{Merlin data}

\begin{figure}[htb!]
	\centering
		\includegraphics[width=0.4\textwidth]{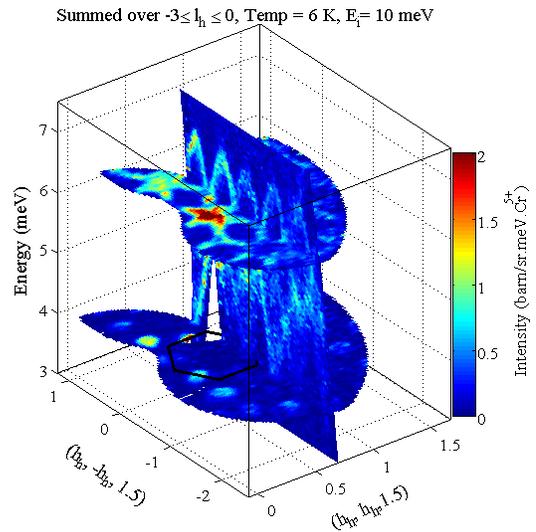}
				\caption{(Color online) Sections of the MERLIN data set displaying the dispersion as a function of energy and wavevector in the hexagonal $(a^{*}_{h},b^{*}_{h})$-plane and integrated over $c^{*}_{h}$, from -3 to 0 r.l.u. The solid black hexagon indicates the hexagonal Brillouin Zone.}
	\label{fig:sliceomatic}
\end{figure}

\begin{figure}[htb!]
	\centering
		\includegraphics[width=0.4\textwidth]{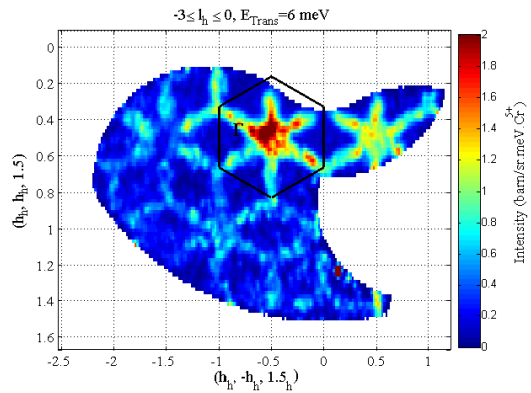}
		\caption{(Color online) Slice showing the scattering for energy transfer $5.9$\,meV as a function of wavevector in the hexagonal $(a^{*}_{h},b^{*}_{h})$-plane, integrated over $c^{*}_{h}$, from -3 to 0 r.l.u.  The solid black hexagon indicates the hexagonal Brillouin Zone.}
	\label{fig:cutMerlin}
\end{figure}

Single crystal inelastic neutron scattering experiments were performed at $6$\,K on the MERLIN spectrometer allowing us to measure the magnetic excitations over a large region of reciprocal space. Figures \ref{fig:sliceomatic} and \ref{fig:cutMerlin} show slices through this dataset where the intensity has been summed over the range $-3<\ell_{h}<0$. Figure \ref{fig:sliceomatic} shows the full extent of the dispersion and figure \ref{fig:cutMerlin} shows the measured intensity at a constant energy transfer of 6\,meV. A complicated excitation spectrum is revealed consisting of three excitation branches. The presence of three modes rather than one confirms the Jahn-Teller distortion proposed by Chapon $et\, al.$ \cite{Chapon} from diffraction measurements. This structural phase transition gives rise to three monoclinic twins each of which produces a separate mode. In agreement with the powder data, the dispersion curves have a gap of $3.4$\,meV and a maximum energy of $6.9$\,meV. The minima and the maxima occur at the midpoints of the faces of the hexagonal Brillouin zone, equivalent to $(1/2_{h},1/2_{h},0_{h})$.

\begin{figure}[htb!]
	\centering
		\includegraphics[width=0.5\textwidth]{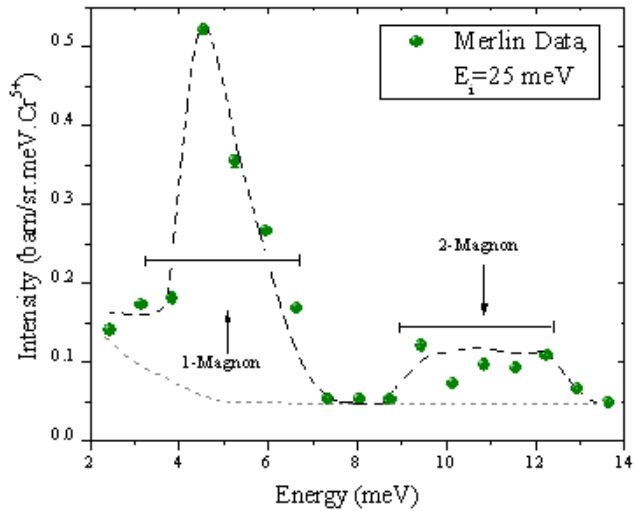}
		\caption{(Color online) $1-$ and $2-$magnon scattering measured at $T=6$\,K with incident neutrons of energy $25$\,meV pointing in the (1$_{h}$,1$_{h}$,0$_{h}$) direction. The data is integrated over the wavevector ranges $-0.2<h_{h}<0.2$, $-1<k_{h}<1$, $-1<\ell_{h}<1$. Lines are guides to the eye.}
	\label{fig:twomagnon}
\end{figure}

In addition to the observed one-magnon scattering, two-magnon scattering is also expected where the neutron creates two magnons simultaneously. Such scattering would be centered at twice the average one-magnon energy ($\approx10.4$\,meV) and with a bandwidth that is the same as the one-magnon bandwidth ($8.7$-$12.1$\,meV). The two magnon cross-section is expected to be considerably weaker than the one-magnon cross-section, but nevertheless observable. Figure \ref{fig:twomagnon} shows the measured scattering intensity as a function of energy transfer integrated over the wavevector range $-0.2<h_{h}<0.2$, $-1<k_{h}<1$, $-1<\ell_{h}<1$. In this region of reciprocal space the intensity of the two-magnon scattering is strongest while that of the one-magnon is weaker. The two magnon peak is clearly observable at $\approx10.5$\,meV. Both one- and two-magnon signal show some structure because of the limited integration range of the data.

\subsubsection{V2-Flex data}

\begin{figure}[htb!]
	\centering
		\includegraphics[width=0.5\textwidth]{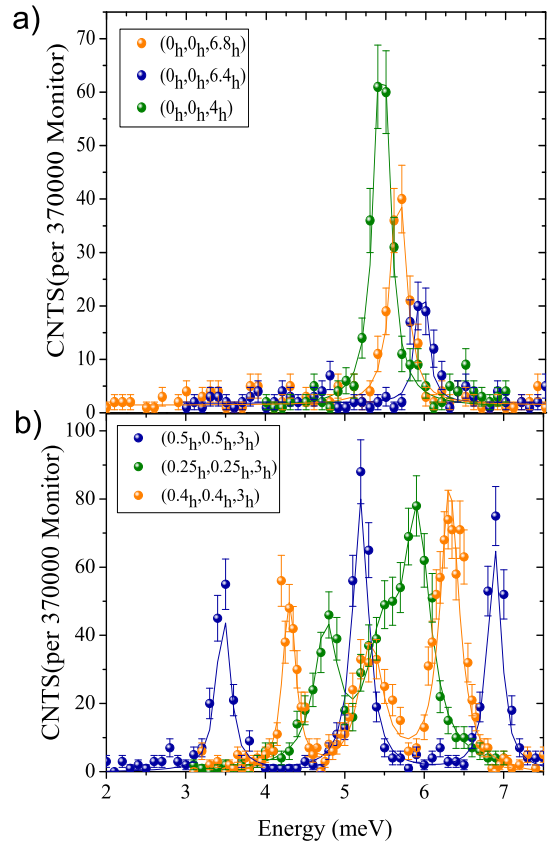}
			\caption{(Color online) Constant-wavevector scans in the (a)$(0_{h},0_{h},\ell_{h})$ and (b)$(h_{h},h_{h},3_{h})$ directions measured at $2$\,K. The solid lines are fits to Lorentzian functions.}
	\label{fig:flex_scans}
\end{figure}

In order to make an accurate determination of the exchange constants, high resolution single crystal inelastic neutron scattering experiments at 2\,K were performed using V2-FLEX. The excitation spectrum was mapped out by scanning the energy transfer at constant wavevector within the $(h_{h},0_{h},\ell_{h})$ and $(h_{h},h_{h},\ell_{h})$ planes. Key scans are shown in Fig. \ref{fig:flex_scans}. As found for the MERLIN data, three peaks were observed at general wavevectors corresponding to the three twinned domains see Fig. \ref{fig:flex_scans}b. These excitations disperse in all directions showing that the interdimer couplings are three-dimensional. The branches merge together in the center of the hexagonal Brillouin zone ($\Gamma$ point), and are degenerate when $h_{h}$ and $k_{h}$ are integer so that a single mode is observed as a function of $\ell_{h}$ see Fig. \ref{fig:flex_scans}a. The bandwidth of the dispersion in this direction gives the interbilayer coupling.
	
\begin{figure}[htb!]
	\centering
		\includegraphics[width=0.5\textwidth]{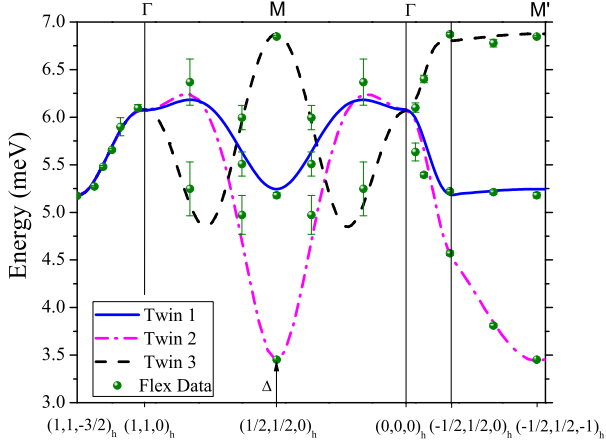}
		\caption{(Color online) Dispersion relation. The blue, magenta and black lines correspond to the fitted dispersion relations of the three monoclinic twins. The green points are the fitted centers of the peaks measured on FLEX. The wavevectors on the x-axis are labeled using the hexagonal notation.}
	\label{fig:dispersion}
\end{figure}

The peaks were fitted to Lorentzians to obtain their positions, (solid lines in Fig. \ref{fig:flex_scans}) and thus map out the dispersion relations of the three twinned domains as a function of energy and wavevector. Figure \ref{fig:dispersion} illustrates the dispersions in various reciprocal lattice directions. The green circles give the excitation energies extracted from each scan and the lines are fits to the RPA model which will be explained in the next section. The three lines (magenta, black and blue) represent the three twins which are rotated by $60$$^\circ$  with respect to each other about the $c_{h}^*$-axis. The dispersion of Twin 1 (blue line) is the same as that of Twin 2 (magenta line) $60$$^\circ$ away in reciprocal space and the same as that of Twin 3 (black line) $120$$^\circ$ away.

\begin{figure}[htb!]
	\centering
			\includegraphics[width=0.5\textwidth]{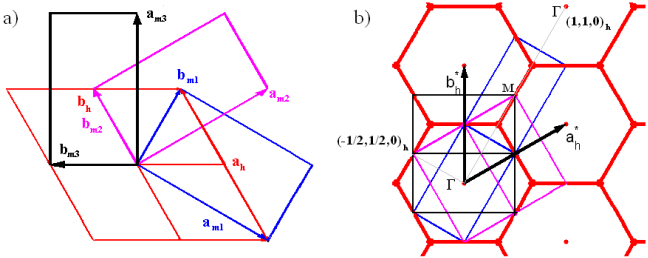}
			\caption{(Color online) (a) Relation between the hexagonal and monoclinic real spaces. Red: hexagonal lattice. Blue: lattice of monoclinic twin 1. Magenta: monoclinic twin 2. Black: monoclinic twin 3. (b) Relationship between the hexagonal and monoclinic reciprocal spaces, projected onto the hexagonal plane, using the same color coding.}
	\label{fig:twins}
\end{figure}

\begin{table*}
\caption{The relationship between the hexagonal lattice parameters and those of the three monoclinic twins is given as well as the relationship between the hexagonal and monoclinic reciprocal spaces.}

\label{table:couplings}
\begin{ruledtabular}
\begin{tabular}{rcrrrrrrrr}
Real Space & & & \\
\hline
Twin1 $(a_{m1},b_{m1}, c_{m1})$ & $a_{m1}=a_{h}-b_{h}$ & $b_{m1}=a_{h}+b_{h}$ & $c_{m1}=a_{h}/3-b_{h}/3+2c_{h}3$ \\
Twin2 $(a_{m2},b_{m2}, c_{m2})$ & $a_{m2}=2a_{h}+b_{h}$ & $b_{m2}=b_{h}$ & $c_{m2}=2a_{h}/3+b_{h}/3+2c_{h}3$ \\
Twin3 $(a_{m3},b_{m3}, c_{m3})$ & $a_{m3}=a_{h}+2b_{h}$ & $b_{m3}=-a_{h}$ & $c_{m3}=a_{h}/3+2b_{h}/3+2c_{h}3$ \\
\hline
Reciprocal Space & & & \\
\hline
Twin1 $(a^{*}_{m1},b^{*}_{m1}, c^{*}_{m1})$ & $a^{*}_{m1}=a^{*}_{h}/2-b^{*}_{h}/2-c^{*}_{h}/2$ & $b^{*}_{m1}=a^{*}_{h}/2+b^{*}_{h}/2$ & $c^{*}_{m1}=3c^{*}_{h}/2$ \\
Twin2 $(a^{*}_{m2},b^{*}_{m2}, c^{*}_{m2})$ & $a^{*}_{m2}=a^{*}_{h}/2-c^{*}_{h}/2$ & $b^{*}_{m2}=-a^{*}_{h}/2+b^{*}_{h}$ & $c^{*}_{m2}=3c^{*}_{h}/2$ \\
Twin3 $(a^{*}_{m3},b^{*}_{m3}, c^{*}_{m3})$ & $a^{*}_{m3}=b^{*}_{h}/2-c^{*}_{h}/2$ & $b^{*}_{m3}=-a^{*}_{h}+b^{*}_{h}/2$ & $c^{*}_{m3}=3c^{*}_{h}/2$ \\
\end{tabular}
\end{ruledtabular}
\end{table*}
The relationship between the hexagonal structure and three monoclinic twins is given in Table 1 where the transformations for both the real and reciprocal lattices are listed. Figure \ref{fig:twins} illustrates these relationships projected onto the `hexagonal' plane, where the original hexagonal lattice is draw in red, and the monoclinic twins are shown in blue, magenta and black. From figure \ref{fig:twins}b it is clear that the corners of the monoclinic Brillouin zones sit at the midpoints of the faces of the hexagonal Brillouin zone ($M$ points). For example, the $M$ point at $(1/2_{h},1/2_{h},0_{h})$ is at the corner of the Brillouin zone of monoclinic Twins 2 and 3, but is at the center of the Brillouin zone of Twin 1. At this wavevector the peak from Twin 2 is at the dispersion minimum while that from Twin 3 is at the dispersion maximum and that from Twin 1 is at the center of the bandwidth, see Fig. \ref{fig:dispersion}. An equivalent way to generate the twins is by permuting each set of exchange constants \{$J_{i}',J_{i}'',J_{i}'''$\}, in this way the dispersion of Twin 1 can be transformed into that of the other two twins.

\begin{figure}[htb!]
	\centering
		\includegraphics[width=0.5\textwidth]{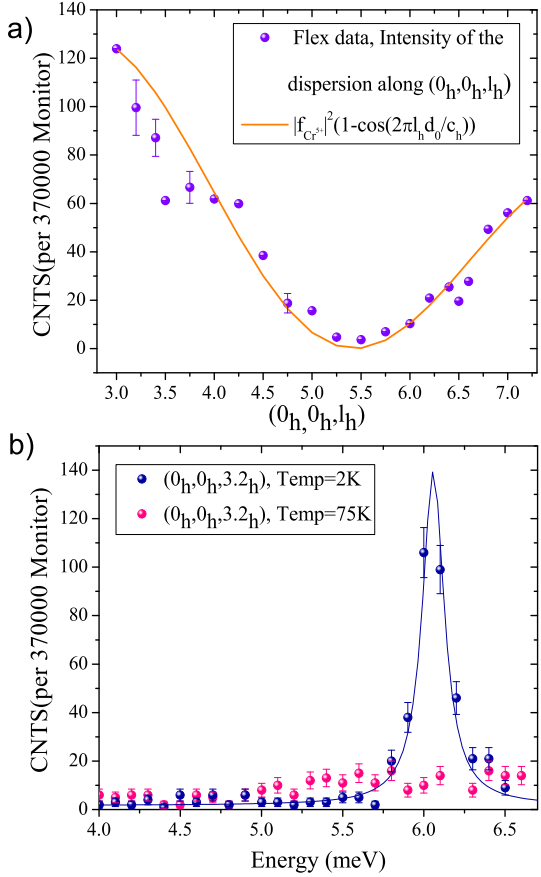}
			\caption{(Color online) (a) Fitted intensity of the magnon in the $(0_{h},0_{h},\ell_{h})$ direction. (b) Comparison of the peak at $\bm{Q}=(0_{h},0_{h},3.2_{h})$  at $2$\,K and $75$\,K.}
	\label{fig:flex_temp_int}
\end{figure}

In a strongly dimerized system such are Sr$_{3}$Cr$_{2}$O$_{8}$ the intensity of the excitations is dominated by the dimer structure factor which is of the form:
\begin{equation}
	\left|f_{\mathrm{Cr}^{5+}}(\left|Q\right|)\right|^{2}(1-cos(2\pi \ell_{h}d_{0}/c_{h})).
\end{equation}
Since the dimer separation $d_{0}$ is parallel to the $c_{h}$-axis we expect an intensity modulation as a function of wavevector transfer along the $\ell_{h}$ axis. The colored symbols in Fig. \ref{fig:flex_temp_int}a show the intensity of the excitations in the $(0_{h},0_{h},\ell_{h})$ direction extracted by fitting the peaks observed in the energy scans (Fig. \ref{fig:flex_scans}a) to Lorentzians. The intensity is found to modulate with a distinctive minimum at $\ell_{h}=5.5$\,r.l.u., and comparison to the theoretical expression (orange line) suggests that the dimer separation is $c_{h}/5.5=3.67$\,\AA\ in good agreement with the nearest neighbor Cr$^{5+}$-Cr$^{5+}$ distance $d_{0}=3.7140(4)$\,\AA. This modulation therefore provides yet further evidence that Sr$_{3}$Cr$_{2}$O$_{8}$ is a system of coupled dimers \cite{Furrer1977}. Finally we performed a higher temperature measurement at $75$\,K. Figure \ref{fig:flex_temp_int}b shows the comparison between the data at $2$\,K and $75$\,K which is well above the temperature of the maximum in the magnetic susceptibility (see Fig. \ref{fig:susceptibility}). At $75$\,K most of the intensity has vanished clarifying the magnetic origin of this excitation.

\section{Analysis}

In order to develop a better understanding of the exchange interactions in Sr$_{3}$Cr$_{2}$O$_{8}$ we compared the extracted dispersion relations to a Random Phase Approximation (RPA) based on the dimer unit. In this model the dispersion relation is \cite{KofuPRLJan}.
\begin{align}
	\hbar\omega\cong\sqrt{J_{0}^{2}+J_{0}\gamma(\mathbf{Q})}\\
		\gamma (\mathbf{Q})=\sum_{i}J(\mathbf{R_{i}})e^{-i\mathbf{Q}\cdot\mathbf{R_{i}}}
\end{align}
where  $\gamma(Q)$ is the Fourier sum of the interdimer interactions. For monoclinic twin $1$,  $\gamma(Q)$ is given by
\begin{align}
\gamma (h_{h}, k_{h}, \ell_{h}) _{Twin1}= 2J'_{1}cos\left(\frac{2}{3}\pi(2h_{h}+k_{h}+\ell_{h})\right) \nonumber \\
+2J''_{1}cos\left(\frac{2}{3}\pi(-h_{h}+k_{h}+\ell_{h})\right) \nonumber \\
+2J'''_{1}cos\left(\frac{2}{3}\pi(-h_{h}-2k_{h}+\ell_{h})\right)\nonumber \\
+2(J'_{2}-J'_{3})cos(2 \pi h_{h})+2(J''_{2}-J''_{3})cos(2 \pi k_{h})\nonumber \\
+2(J'''_{2}-J'''_{3})cos(2 \pi(h_{h}+k_{h})) \nonumber \\
+2J'_{4}cos\left(\frac{2}{3}\pi(2h_{h}+4k_{h}+\ell_{h})\right) \nonumber \\
+2J''_{4}cos\left(\frac{2}{3}\pi(2h_{h}-2k_{h}+\ell_{h})\right)\nonumber \\
+2J'''_{4}cos\left(\frac{2}{3}\pi(-4h_{h}-2k_{h}+\ell_{h})\right),
\end{align}
where we have included the set of $4^{th}$ order interactions $J_{4}'$, $J_{4}''$, $J_{4}'''$ which were found to be significant in the analysis of Ba$_{3}$Mn$_{2}$O$_{8}$. Note that in this model it is not possible to distinguish between the sets $J_{2}'$, $J_{2}''$, $J_{2}'''$ and $J_{3}'$, $J_{3}''$, $J_{3}'''$ and only the difference between these exchange parameters matters i.e. $J_{2}'-J_{3}'$ etc. Also note that the minus sign expresses the fact that $J_{2}'$ and $J_{3}'$ compete if they have the same sign (e.g. both antiferromagnetic) but reinforce each other if they have different signs. $\gamma(Q)$ for the other twins can be generated from that of twin $1$ by permuting the exchange constants as follows
\begin{align}
\left\{J_{1,2-3,4}',J_{1,2-3,4}'',J_{1,2-3,4}'''\right\}_{\mathrm{Twin1}} \nonumber \\ \rightarrow \left\{J_{1,2-3,4}'',J_{1,2-3,4}''',J_{1,2-3,4}'\right\}_{\mathrm{Twin2}} \nonumber \\ \rightarrow \left\{J_{1,2-3,4}''',J_{1,2-3,4}',J_{1,2-3,4}''\right\}_{\mathrm{Twin3}} \nonumber
\end{align}
In this way the magnetic excitations for all three twins can be calculated. The RPA model was compared to the dispersion relations extracted from the V2-FLEX data (see Fig. \ref{fig:dispersion}). In these experiments the excitations were measured in two planes the $(h_{h},0_{h},\ell_{h})$ and $(h_{h},h_{h},\ell_{h})$ and for these specific planes equation 13 reduces to:
\begin{align}
\gamma (h_{h}, 0_{h}, \ell_{h})_{Twin1} =2J'_{1}cos\left(\frac{2}{3}\pi(2h_{h}+\ell_{h})\right)\nonumber \\
+(2J''_{1}+2J'''_{1})cos\left(\frac{2}{3}\pi(-h_{h}+\ell_{h})\right) \nonumber \\
+(2(J'_{2}-J'_{3})+2(J'''_{2}-J'''_{3}))cos(2\pi h_{h})+2(J''_{2}-J''_{3})\nonumber \\
+(2J'_{4}+2J''_{4})cos\left(\frac{2}{3}\pi(2h_{h}+\ell_{h})\right) \nonumber \\
+ 2J'''_{4}cos\left(\frac{2}{3}\pi(-4h_{h}+\ell_{h})\right)
\end{align}
and
\begin{align}
\gamma (h_{h}, h_{h}, \ell_{h})_{Twin1} =2J'_{1}cos\left(\frac{2}{3}\pi(3h_{h}+\ell_{h})\right) \nonumber \\
+2J''_{1}cos\left(\frac{2}{3}\pi(\ell_{h})\right)+2J'''_{1}cos\left(\frac{2}{3}\pi(-3h_{h}+\ell_{h})\right) \nonumber \\
+2((J'_{2}-J'_{3})+2(J''_{2}-J''_{3}))cos(2\pi h_{h}) \nonumber \\
+2(J'''_{2}-J'''_{3})cos(4\pi h_{h})+2J'_{4}cos\left(\frac{2}{3}\pi(6h_{h}+\ell_{h})\right) \nonumber \\
+2J''_{4}cos\left(\frac{2}{3}\pi \ell_{h}\right)+2J'''_{4}cos\left(\frac{2}{3}\pi(-6h_{h}+\ell_{h})\right).
\end{align}

These equations can be simplified further if scans along key directions are considered, e.g. for directions where $h_{h}$ and $k_{h}$ are integers and $\ell_{h}$ varies, the set of interactions $J_{2}'-J_{3}'$, $J_{2}''-J_{3}''$,  $J_{2}'''-J_{3}'''$ do not modulate the structure but only give an overall constant, thus allowing us to obtain the interactions $J_{0}$, $J_{1}'$, $J_{1}''$, $J_{1}'''$ and $J_{4}'$, $J_{4}''$, $J_{4}'''$. The directions with $h_{h}$ and $k_{h}$ half-integer were then used to get the values for $J_{2}'-J_{3}'$, $J_{2}''-J_{3}''$,  $J_{2}'''-J_{3}'''$. The fitted exchange constants are listed in Table 2 and the resulting dispersion relations for the three monoclinic twins are illustrated in figure \ref{fig:dispersion} and clearly show that this model provides a good approximation to the data.
\begin{table*}
\caption{The exchange constants in units of meV extracted from the RPA model, compared to the values for Ba$_{3}$Cr$_{2}$O$_{8}$ and Ba$_{3}$Mn$_{2}$O$_{8}$. Antiferromagnetic interactions are positive while ferromagnetic interactions are negative.} \label{table:couplings}
\begin{ruledtabular}
\begin{tabular}{rcrrrrrrrr}
Exchange Constants & Sr$_{3}$Cr$_{2}$O$_{8}$ & Ba$_{3}$Cr$_{2}$O$_{8}$ \cite{KofuPRLJan} & Ba$_{3}$Mn$_{2}$O$_{8}$\cite{StonePRL100}\\
\hline
$J_{0}$ & 5.551(9)&2.38 &1.642 \\
$J_{1}'$ & -0.04(1)&-0.15 &-0.118\\
$J_{1}''$& 0.24(1)&0.08 &-0.118\\
$J_{1}'''$& 0.25(1)& 0.10 & -0.118\\
$J_{2}'-J_{3}'$& 0.751(9)&0.10 & 0.256-0.142\\
$J_{2}''-J_{3}''$& -0.543(9)&-0.52 & 0.256-0.142\\
$J_{2}'''-J_{3}'''$&-0.120(9) &0.07 &0.256-0.142\\
$J_{4}'$& 0.10(2)&0.10 & -0.037\\
$J_{4}''$& -0.05(1)&0.04 &-0.037\\
$J_{4}'''$& 0.04(1)&0.09 &-0.037\\
$J'=\left|J'_{1}\right|+\left|J''_{1}\right|+\left|J'''_{1}\right|+\left|J'_{4}\right|+\left|J''_{4}\right|+\left|J'''_{4}\right|$ &$ J'=3.6(1)$&$ J'=1.94$&$ J'=1.147$\\
$+2(\left|J'_{2}-J'_{3}\right|)+2(\left|J''_{2}-J''_{3}\right|)+2(\left|J'''_{2}-J'''_{3}\right|)$ &$J'/J_{0}=0.6455$&$J'/J_{0}=0.8151$ & $J'/J_{0}=0.6983$\\
\end{tabular}
\end{ruledtabular}
\end{table*}

\begin{figure}[htb!]
	\centering
		\includegraphics[width=0.5\textwidth]{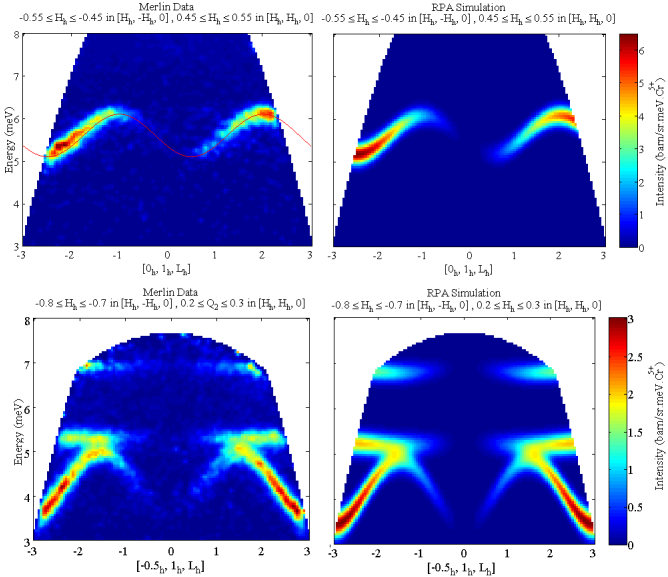}
		\caption{(Color online) Shows slices of the Merlin Data (left panels) and the RPA simulation (right panels) for the directions $(0_h,1_h,\ell_h)$ and $(-0.5_h,1_h,\ell_h)$, with E$_i=$10\,meV and 6\,K.}
	\label{fig:RPAsimulation}
\end{figure}

The neutron scattering cross-section from these modes can also be calculated in the RPA model and is given by
\begin{equation}
\frac{d^{2}\sigma(\mathbf{Q}, E)}{d\Omega dE}\approx \frac{\left|f_{\mathrm{Cr}^{5+}}(\left|Q\right|)\right|^{2}(1-cos(\frac{2\pi \ell_{h}d_{0}}{c_{h}}))e^{(-(E-\hbar \omega)^{2}/\Delta E^{2})}}{\hbar \omega (1-e^{E/k_{B}T})}
\end{equation}
Where $E$ is the energy transfer, $\Delta E$ is the energy resolution of the instrument which is assumed to be Gaussian and $T$ is the temperature. The cosine term comes from the structure factor of a single dimer (see equation (10)). Using the fitted values of the exchange constants (from Table 2), a simulation was performed of the MERLIN experiment. Figure \ref{fig:RPAsimulation} shows good agreement between the data and the simulation for various slices. The RPA model along with the fitted exchange constants was also used to reproduce the data from the inelastic powder measurement performed on V3-NEAT as shown in Fig. \ref{fig:powder} (right panel). Cuts as a function of energy and wavevector were performed on this simulation and compare well to the data see figures \ref{fig:bandwidth} and \ref{fig:firstmoment}. Finally, the fitted interactions were used to calculate the effective field $J'=3.583$\,meV (see Table 2) which was used along with the fitted value of $J_{0}$ to simulate our susceptibility data using equation 3. This simulation is shown in figure \ref{fig:susceptibility} and is consistent with the data.

\section{Discussion}

Our DC susceptibility, high field magnetization and inelastic neutron scattering results provide a consistent picture of Sr$_{3}$Cr$_{2}$O$_{8}$ as a dimerized spin-$1/2$ antiferromagnet. The intradimer exchange constant is identified as the intrabilayer interaction $J_{0}$ which has a value of $5.55$\,meV as found from both susceptibility and inelastic neutron scattering. This dimerization gives rise to a singlet ground state and gapped one-magnon excitations. There are in addition significant interdimer interactions that allow the dimer excitations to hop and develop a dispersion. The dispersion of the one-magnon mode produces a bandwidth extending between the gap energy of $3.5$\,meV and the maximum value of  $7.0$\,meV as found by magnetization and inelastic neutron scattering. A second band due to two-magnon excitations was also found centered at $10.5$\,meV. The inelastic neutron scattering data was fitted to a RPA model based on the dimer unit and the intradimer and interdimer interactions were determined.

Our results can be compared to previous experimental work on Sr$_{3}$Cr$_{2}$O$_{8}$. Y. Singh and D.C. Johnston \cite{Singh} measured the DC susceptibility and obtained a value of $J_{0} = 5.17(1)$\,meV in good agreement with our value. In contrast their value for the sum of the interdimer exchange constants is $J'=$0.5(2)\,meV, which, while consistent with our best fit value from susceptibility, is much smaller than our more accurate value of $J'=3.6(1)$\,meV obtained from magnetization and inelastic neutron scattering. Inspection of the fitting process shows that the susceptibility data can however be well modeled by a wide range of $J'$ values including $J'=3.6(1)$\,meV.

Chapon $et\, al.$ \cite{Chapon} show from diffraction measurements that Sr$_{3}$Cr$_{2}$O$_{8}$ undergoes a Jahn-Teller distortion resulting in a lowering of the crystal symmetry to monoclinic. This distortion would produce spatially anisotropic exchange interactions and result in three twinned domains of equal volume below the phase transition. Our results are consistent with this distortion since we observe three excitation modes arising from the three twins and our fitted exchange interactions are indeed spatially anisotropic. The exchange interactions in Sr$_{3}$Cr$_{2}$O$_{8}$ have been calculated theoretically using the Extended H\"{u}ckel Tight Binding Model \cite{Chapon} and the theoretical value of $J'/J_{0}$ is $0.11$, which is much smaller than our experimentally determined value for this ratio of $0.6455$. Finally, Chapon $et\, al.$ used powder inelastic neutron scattering measurements to show that the intradimer interaction is the intrabilayer interaction with a value of $J_{0}=5.46$\,meV in good agreement with our value, however the size of the interdimer interactions could not be obtain from the bandwidth of the excitations due to the low resolution of the measurement.

Our results can be compared to those for the related compounds Ba$_{3}$Mn$_{2}$O$_{8}$ and Ba$_{3}$Cr$_{2}$O$_{8}$. The magnetic ion Mn$^{5+}$ has spin-$1$ compared to spin-$1/2$ for Cr$^{5+}$, and is not Jahn-Teller active, as a result Ba$_{3}$Mn$_{2}$O$_{8}$ retains hexagonal symmetry down to low temperatures and the interactions remain frustrated. A single one-magnon excitation branch is observed with minima at $(1/3_{h},2/3_{h},1/2_{h})$  and equivalent positions \cite{StonePRL100}. In contrast Ba$_{3}$Cr$_{2}$O$_{8}$ like Sr$_{3}$Cr$_{2}$O$_{8}$ undergoes a Jahn-Teller distortion to monoclinc symmetry leading to spatially anisotropic exchange interactions and three excitation branches arising from the three twinned domains with minima at the $(1/2_{h},1_{h},0_{h})$ etc positions.

The values of intradimer interaction $J_{0}$ are smaller in the Ba compounds being $1.64$\,meV for Ba$_{3}$Mn$_{2}$O$_{8}$ \cite{StonePRL100} and $2.38$\,meV for Ba$_{3}$Cr$_{2}$O$_{8}$ \cite{KofuPRLJan}. This may simply be due to the intradimer distance which is greater for Ba$_{3}$Cr$_{2}$O$_{8}$ ($3.934$\,\AA) and Ba$_{3}$Mn$_{2}$O$_{8}$ ($3.984$\,\AA) than for Sr$_{3}$Cr$_{2}$O$_{8}$ ($3.842$\,\AA) allowing better overlap of orbitals. The sizes of the interdimer exchange constants are also on average smaller in these compounds although the ratio of intradimer to interdimer couplings is larger. In Ba$_{3}$Cr$_{2}$O$_{8}$ $J'=1.94$\,meV, and $J'/J_{0}= 0.8151$ while for Ba$_{3}$Mn$_{2}$O$_{8}$ $J'/J_{0}=0.6983$ showing that the ratio of excitation bandwidth to the average mode energy is greater than in Sr$_{3}$Cr$_{2}$O$_{8}$ where $J'/J_{0}= 0.6455$. Finally the spin gap of Sr$_{3}$Cr$_{2}$O$_{8}$ is $3.5$\,meV and is thus significantly larger than that of Ba$_{3}$Mn$_{2}$O$_{8}$ ($1.08$\,meV) and Ba$_{3}$Cr$_{2}$O$_{8}$($1.38$\,meV).
	
Plenty of additional work on Sr$_{3}$Cr$_{2}$O$_{8}$ is needed to fully understand this system. Optical absorption measurements are currently underway to measure the orbital energy level diagram and confirm the orbital ordering proposed by Chapon $et\, al.$ \cite{Chapon}. Electron Parametnetic resonance (EPR) is also taking place to determine the $g$-tensor and to search for higher order terms in the Hamiltonian e.g. Dzyaloshinskii-Moriya interactions \cite{KofuPRLMay}. We also plan Raman scattering measurements to further investigate the magnetic excitation spectrum and in particular the multi-magnon features, and magnetostriction measurements to study the magneto-elastic coupling. Finally Sr$_{3}$Cr$_{2}$O$_{8}$ is a candidate compound for Bose-Einstein condensation of magnons, where an applied magnetic field of $H_{c}=30.9(4)$\,T is required to close the spin gap and start condensing magnons in the ground state, while full saturation is achieved at a field of $H_{s}=61.9(3)$\,T. Very recently an interesting new paper was reported on the LANL preprint server \cite{AczelLANL} describing heat capacity and the magnetocaloric effect measurements performed at high magnetic fields to map out the phase diagram as a function of field and temperature. The extracted critical exponent of the ordering temperature as a function of reduced field around $H_{c}$ is indeed in agreement with the universality class predicted for a three-dimensional Bose-Einstein Condensate and measurements of the critical exponents at the upper critical field are in progress.

\begin{acknowledgments}
We acknowledge Y. Singh, D. C. Johnston for their help with the preparation for powder samples and D.N. Argyriou for the use of the Crystal lab at HZB. R.A. Ewings, T.G. Perring provided the Horace program along with their expertise for viewing and analyzing the Merlin data.
\end{acknowledgments}


\begin{thebibliography}{23}
\expandafter\ifx\csname natexlab\endcsname\relax\def\natexlab#1{#1}\fi
\expandafter\ifx\csname bibnamefont\endcsname\relax
  \def\bibnamefont#1{#1}\fi
\expandafter\ifx\csname bibfnamefont\endcsname\relax
  \def\bibfnamefont#1{#1}\fi
\expandafter\ifx\csname citenamefont\endcsname\relax
  \def\citenamefont#1{#1}\fi
\expandafter\ifx\csname url\endcsname\relax
  \def\url#1{\texttt{#1}}\fi
\expandafter\ifx\csname urlprefix\endcsname\relax\def\urlprefix{URL }\fi
\providecommand{\bibinfo}[2]{#2}
\providecommand{\eprint}[2][]{\url{#2}}

\bibitem[{\citenamefont{Ruegg et~al.}(2003)\citenamefont{Ruegg, Cavadini,
  Furrer, Gudel, Kramer, Mutka, Habicht, Vorderwisch, and
  Wildes}}]{Rueggnature}
\bibinfo{author}{\bibfnamefont{C.}~\bibnamefont{Ruegg}},
  \bibinfo{author}{\bibfnamefont{N.}~\bibnamefont{Cavadini}},
  \bibinfo{author}{\bibfnamefont{A.}~\bibnamefont{Furrer}},
  \bibinfo{author}{\bibfnamefont{H.~U.} \bibnamefont{Gudel}},
  \bibinfo{author}{\bibfnamefont{K.}~\bibnamefont{Kramer}},
  \bibinfo{author}{\bibfnamefont{H.}~\bibnamefont{Mutka}},
  \bibinfo{author}{\bibfnamefont{A.~K.} \bibnamefont{Habicht}},
  \bibinfo{author}{\bibfnamefont{P.}~\bibnamefont{Vorderwisch}},
  \bibnamefont{and} \bibinfo{author}{\bibfnamefont{A.}~\bibnamefont{Wildes}},
  \bibinfo{journal}{Nature(London)} \textbf{\bibinfo{volume}{423}},
  \bibinfo{pages}{62} (\bibinfo{year}{2003}).

\bibitem[{\citenamefont{Nikuni et~al.}(2000)\citenamefont{Nikuni, Oshikawa,
  Oosawa, and Tanaka}}]{NikuniPRL84}
\bibinfo{author}{\bibfnamefont{T.}~\bibnamefont{Nikuni}},
  \bibinfo{author}{\bibfnamefont{M.}~\bibnamefont{Oshikawa}},
  \bibinfo{author}{\bibfnamefont{A.}~\bibnamefont{Oosawa}}, \bibnamefont{and}
  \bibinfo{author}{\bibfnamefont{H.}~\bibnamefont{Tanaka}},
  \bibinfo{journal}{Phys. Rev. Lett.} \textbf{\bibinfo{volume}{84}},
  \bibinfo{pages}{5868} (\bibinfo{year}{2000}).

\bibitem[{\citenamefont{Sasago et~al.}(1997)\citenamefont{Sasago, Uchinokura,
  Zheludev, and Shirane}}]{SasagoPRB55}
\bibinfo{author}{\bibfnamefont{Y.}~\bibnamefont{Sasago}},
  \bibinfo{author}{\bibfnamefont{K.}~\bibnamefont{Uchinokura}},
  \bibinfo{author}{\bibfnamefont{A.}~\bibnamefont{Zheludev}}, \bibnamefont{and}
  \bibinfo{author}{\bibfnamefont{G.}~\bibnamefont{Shirane}},
  \bibinfo{journal}{Phys. Rev. B} \textbf{\bibinfo{volume}{55}},
  \bibinfo{pages}{8357} (\bibinfo{year}{1997}).

\bibitem[{\citenamefont{Ruegg et~al.}(2007)\citenamefont{Ruegg, McMorrow,
  Normand, Ronnow, Sebastian, Fisher, Batista, Gvasaliya, Niedermayer, and
  Stahn}}]{RueggPRL98}
\bibinfo{author}{\bibfnamefont{C.}~\bibnamefont{Ruegg}},
  \bibinfo{author}{\bibfnamefont{D.~F.} \bibnamefont{McMorrow}},
  \bibinfo{author}{\bibfnamefont{B.}~\bibnamefont{Normand}},
  \bibinfo{author}{\bibfnamefont{H.~M.} \bibnamefont{Ronnow}},
  \bibinfo{author}{\bibfnamefont{S.~E.} \bibnamefont{Sebastian}},
  \bibinfo{author}{\bibfnamefont{I.~R.} \bibnamefont{Fisher}},
  \bibinfo{author}{\bibfnamefont{C.~D.} \bibnamefont{Batista}},
  \bibinfo{author}{\bibfnamefont{S.~N.} \bibnamefont{Gvasaliya}},
  \bibinfo{author}{\bibfnamefont{C.}~\bibnamefont{Niedermayer}},
  \bibnamefont{and} \bibinfo{author}{\bibfnamefont{J.}~\bibnamefont{Stahn}},
  \bibinfo{journal}{Phys. Rev. Lett.} \textbf{\bibinfo{volume}{98}},
  \bibinfo{pages}{017202} (\bibinfo{year}{2007}).

\bibitem[{\citenamefont{Ruegg et~al.}(2004)\citenamefont{Ruegg, Furrer,
  Sheptyakov, Strassle, Kramer, Gudel, and Melesi}}]{RueggPRL93}
\bibinfo{author}{\bibfnamefont{C.}~\bibnamefont{Ruegg}},
  \bibinfo{author}{\bibfnamefont{A.}~\bibnamefont{Furrer}},
  \bibinfo{author}{\bibfnamefont{D.}~\bibnamefont{Sheptyakov}},
  \bibinfo{author}{\bibfnamefont{T.}~\bibnamefont{Strassle}},
  \bibinfo{author}{\bibfnamefont{K.~W.} \bibnamefont{Kramer}},
  \bibinfo{author}{\bibfnamefont{H.~U.} \bibnamefont{Gudel}}, \bibnamefont{and}
  \bibinfo{author}{\bibfnamefont{L.}~\bibnamefont{Melesi}}, \bibinfo{journal}{Phys.
  Rev. Lett.} \textbf{\bibinfo{volume}{93}}, \bibinfo{pages}{257201}
  (\bibinfo{year}{2004}).

\bibitem[{\citenamefont{Kageyama et~al.}(1999)\citenamefont{Kageyama,
  Yoshimura, Stern, Mushnikov, Onizuka, Kato, Kosuge, Slichter, Goto, and
  Ueda}}]{KageyamaPRL82}
\bibinfo{author}{\bibfnamefont{H.}~\bibnamefont{Kageyama}},
  \bibinfo{author}{\bibfnamefont{K.}~\bibnamefont{Yoshimura}},
  \bibinfo{author}{\bibfnamefont{R.}~\bibnamefont{Stern}},
  \bibinfo{author}{\bibfnamefont{N.~V.} \bibnamefont{Mushnikov}},
  \bibinfo{author}{\bibfnamefont{K.}~\bibnamefont{Onizuka}},
  \bibinfo{author}{\bibfnamefont{M.}~\bibnamefont{Kato}},
  \bibinfo{author}{\bibfnamefont{K.}~\bibnamefont{Kosuge}},
  \bibinfo{author}{\bibfnamefont{C.~P.}~\bibnamefont{Slichter}},
  \bibinfo{author}{\bibfnamefont{T.}~\bibnamefont{Goto}}, \bibnamefont{and}
  \bibinfo{author}{\bibfnamefont{Y.}~\bibnamefont{Ueda}},
  \bibinfo{journal}{Phys. Rev. Lett.} \textbf{\bibinfo{volume}{82}},
  \bibinfo{pages}{3168} (\bibinfo{year}{1999}).

\bibitem[{\citenamefont{Stone et~al.}(2008)\citenamefont{Stone, Lumsden, Chang,
  Samulon, Batista, and Fisher}}]{StonePRL100}
\bibinfo{author}{\bibfnamefont{M.~B.} \bibnamefont{Stone}},
  \bibinfo{author}{\bibfnamefont{M.~D.} \bibnamefont{Lumsden}},
  \bibinfo{author}{\bibfnamefont{S.}~\bibnamefont{Chang}},
  \bibinfo{author}{\bibfnamefont{E.~C.} \bibnamefont{Samulon}},
  \bibinfo{author}{\bibfnamefont{C.~D.} \bibnamefont{Batista}},
  \bibnamefont{and} \bibinfo{author}{\bibfnamefont{I.~R.}
  \bibnamefont{Fisher}}, \bibinfo{journal}{Phys. Rev. Lett.}
  \textbf{\bibinfo{volume}{100}}, \bibinfo{pages}{237201}
  (\bibinfo{year}{2008}).

\bibitem[{\citenamefont{Koo et~al.}(2006)\citenamefont{Koo, Lee, and
  Whangbo}}]{Koo}
\bibinfo{author}{\bibfnamefont{H.-J.} \bibnamefont{Koo}},
  \bibinfo{author}{\bibfnamefont{K.-S.} \bibnamefont{Lee}}, \bibnamefont{and}
  \bibinfo{author}{\bibfnamefont{M.-H.} \bibnamefont{Whangbo}},
  \bibinfo{journal}{Inorg. Chem.} \textbf{\bibinfo{volume}{45}},
  \bibinfo{pages}{10743} (\bibinfo{year}{2006}).

\bibitem[{\citenamefont{K. et~al.}(2009)\citenamefont{K., Al-Hassanieh,
  Samulon, Brooks, Clark, Kuhns, Lumata, Reyes, Fisher, Brown
  et~al.}}]{Suharxiv}
\bibinfo{author}{\bibfnamefont{S.~S.} \bibnamefont{K.}},
  \bibinfo{author}{\bibfnamefont{A.}~\bibnamefont{Al-Hassanieh}},
  \bibinfo{author}{\bibfnamefont{E.~C.} \bibnamefont{Samulon}},
  \bibinfo{author}{\bibfnamefont{J.~S.} \bibnamefont{Brooks}},
  \bibinfo{author}{\bibfnamefont{W.~G.} \bibnamefont{Clark}},
  \bibinfo{author}{\bibfnamefont{P.~L.} \bibnamefont{Kuhns}},
  \bibinfo{author}{\bibfnamefont{L.~L.} \bibnamefont{Lumata}},
  \bibinfo{author}{\bibfnamefont{A.}~\bibnamefont{Reyes}},
  \bibinfo{author}{\bibfnamefont{I.~R.} \bibnamefont{Fisher}},
  \bibinfo{author}{\bibfnamefont{S.~E.} \bibnamefont{Brown}},
  \bibnamefont{et~al.}, \bibinfo{journal}{arXiv:0905.0718v1 [cond-mat.str-el]}
  (\bibinfo{year}{2009}).

\bibitem[{\citenamefont{Kofu et~al.}(2009{\natexlab{a}})\citenamefont{Kofu,
  Kim, Ji, Lee, Ueda, Qiu, Kang, Green, and Ueda}}]{KofuPRLJan}
\bibinfo{author}{\bibfnamefont{M.}~\bibnamefont{Kofu}},
  \bibinfo{author}{\bibfnamefont{J.~H.} \bibnamefont{Kim}},
  \bibinfo{author}{\bibfnamefont{S.}~\bibnamefont{Ji}},
  \bibinfo{author}{\bibfnamefont{S.~H.} \bibnamefont{Lee}},
  \bibinfo{author}{\bibfnamefont{H.}~\bibnamefont{Ueda}},
  \bibinfo{author}{\bibfnamefont{Y.}~\bibnamefont{Qiu}},
  \bibinfo{author}{\bibfnamefont{H.~J.} \bibnamefont{Kang}},
  \bibinfo{author}{\bibfnamefont{M.~A.} \bibnamefont{Green}}, \bibnamefont{and}
  \bibinfo{author}{\bibfnamefont{Y.}~\bibnamefont{Ueda}},
  \bibinfo{journal}{Phys. Rev. Lett.} \textbf{\bibinfo{volume}{102}},
  \bibinfo{pages}{037206} (\bibinfo{year}{2009}{\natexlab{a}}).

\bibitem[{\citenamefont{Kofu et~al.}(2009{\natexlab{b}})\citenamefont{Kofu,
  Ueda, Nojiri, Oshima, Zenmoto, Rule, Gerischer, Lake, Batista, Ueda
  et~al.}}]{KofuPRLMay}
\bibinfo{author}{\bibfnamefont{M.}~\bibnamefont{Kofu}},
  \bibinfo{author}{\bibfnamefont{H.}~\bibnamefont{Ueda}},
  \bibinfo{author}{\bibfnamefont{H.}~\bibnamefont{Nojiri}},
  \bibinfo{author}{\bibfnamefont{Y.}~\bibnamefont{Oshima}},
  \bibinfo{author}{\bibfnamefont{T.}~\bibnamefont{Zenmoto}},
  \bibinfo{author}{\bibfnamefont{K.~C.} \bibnamefont{Rule}},
  \bibinfo{author}{\bibfnamefont{S.}~\bibnamefont{Gerischer}},
  \bibinfo{author}{\bibfnamefont{B.}~\bibnamefont{Lake}},
  \bibinfo{author}{\bibfnamefont{C.~D.} \bibnamefont{Batista}},
  \bibinfo{author}{\bibfnamefont{Y.}~\bibnamefont{Ueda}}, \bibnamefont{et~al.},
  \bibinfo{journal}{Phys. Rev. Lett.} \textbf{\bibinfo{volume}{102}},
  \bibinfo{pages}{177204} (\bibinfo{year}{2009}{\natexlab{b}}).

\bibitem[{\citenamefont{Aczel et~al.}(2009)\citenamefont{Aczel, Kohama, Jaime,
  Ninios, Chan, Balicas, Dabkowska, and Luke}}]{AczelPRB2009}
\bibinfo{author}{\bibfnamefont{A.~A.} \bibnamefont{Aczel}},
  \bibinfo{author}{\bibfnamefont{Y.}~\bibnamefont{Kohama}},
  \bibinfo{author}{\bibfnamefont{M.}~\bibnamefont{Jaime}},
  \bibinfo{author}{\bibfnamefont{K.}~\bibnamefont{Ninios}},
  \bibinfo{author}{\bibfnamefont{H.~B.} \bibnamefont{Chan}},
  \bibinfo{author}{\bibfnamefont{L.}~\bibnamefont{Balicas}},
  \bibinfo{author}{\bibfnamefont{H.~A.} \bibnamefont{Dabkowska}},
  \bibnamefont{and} \bibinfo{author}{\bibfnamefont{G.~M.} \bibnamefont{Luke}},
  \bibinfo{journal}{Phys. Rev. B} \textbf{\bibinfo{volume}{79}},
  \bibinfo{pages}{100409(R)} (\bibinfo{year}{2009}).

\bibitem[{\citenamefont{Aczel et~al.}(2009{\natexlab{b}})\citenamefont{Aczel,
  Kohama, Marcenat, Weickert, Jaime, McDonald, Selesnic, Dabkowska, and
  Luke}}]{AczelLANL}
\bibinfo{author}{\bibfnamefont{A.~A.} \bibnamefont{Aczel}},
  \bibinfo{author}{\bibfnamefont{Y.}~\bibnamefont{Kohama}},
  \bibinfo{author}{\bibfnamefont{C.}~\bibnamefont{Marcenat}},
  \bibinfo{author}{\bibfnamefont{F.}~\bibnamefont{Weickert}},
  \bibinfo{author}{\bibfnamefont{M.}~\bibnamefont{Jaime}},
  \bibinfo{author}{\bibfnamefont{R.~D.} \bibnamefont{McDonald}},
  \bibinfo{author}{\bibfnamefont{S.~D.} \bibnamefont{Selesnic}},
  \bibinfo{author}{\bibfnamefont{H.~A.} \bibnamefont{Dabkowska}},
  \bibnamefont{and} \bibinfo{author}{\bibfnamefont{G.~M.} \bibnamefont{Luke}},
  \bibinfo{journal}{arXiv:0908.3049 [cond-mat.mtrl-sci]}
  (\bibinfo{year}{2009}{\natexlab{b}}).

\bibitem[{\citenamefont{Cuno and Mullerbuschbaum}(1989)}]{Cuno1989}
\bibinfo{author}{\bibfnamefont{E.}~\bibnamefont{Cuno}} \bibnamefont{and}
  \bibinfo{author}{\bibfnamefont{H.}~\bibnamefont{Mullerbuschbaum}},
  \bibinfo{journal}{Z. Anorg. Allg. Chem.} \textbf{\bibinfo{volume}{572}},
  \bibinfo{pages}{95} (\bibinfo{year}{1989}).

\bibitem[{\citenamefont{Jacob and Abraham}(2000)}]{Jacob2000}
\bibinfo{author}{\bibfnamefont{K.}~\bibnamefont{Jacob}} \bibnamefont{and}
  \bibinfo{author}{\bibfnamefont{K.}~\bibnamefont{Abraham}},
  \bibinfo{journal}{J. Phase Equilibria} \textbf{\bibinfo{volume}{21}},
  \bibinfo{pages}{46} (\bibinfo{year}{2000}).

\bibitem[{\citenamefont{Singh and Johnston}(2007)}]{Singh}
\bibinfo{author}{\bibfnamefont{Y.}~\bibnamefont{Singh}} \bibnamefont{and}
  \bibinfo{author}{\bibfnamefont{D.~C.} \bibnamefont{Johnston}},
  \bibinfo{journal}{Phys. Rev. B} \textbf{\bibinfo{volume}{76}},
  \bibinfo{pages}{012407} (\bibinfo{year}{2007}).

\bibitem[{\citenamefont{Chapon et~al.}(2008)\citenamefont{Chapon, Stock,
  Radaelli, and Martin}}]{Chapon}
\bibinfo{author}{\bibfnamefont{L.}~\bibnamefont{Chapon}},
  \bibinfo{author}{\bibfnamefont{C.}~\bibnamefont{Stock}},
  \bibinfo{author}{\bibfnamefont{P.}~\bibnamefont{Radaelli}}, \bibnamefont{and}
  \bibinfo{author}{\bibfnamefont{C.}~\bibnamefont{Martin}},
  \bibinfo{journal}{arXiv:0807.0877v2 [cond-mat.mtrl-sci]}
  (\bibinfo{year}{2008}).

\bibitem[{\citenamefont{Islam et~al.}(2009)\citenamefont{Islam,
  Quintero-Castro, Lake, Siemensmeyer, Kiefer, Skourski, and
  Hermannsdorfer}}]{Nazmul}
\bibinfo{author}{\bibfnamefont{A.~T. M.~N.} \bibnamefont{Islam}},
  \bibinfo{author}{\bibfnamefont{D.}~\bibnamefont{Quintero-Castro}},
  \bibinfo{author}{\bibfnamefont{B.}~\bibnamefont{Lake}},
  \bibinfo{author}{\bibfnamefont{K.}~\bibnamefont{Siemensmeyer}},
  \bibinfo{author}{\bibfnamefont{K.}~\bibnamefont{Kiefer}},
  \bibinfo{author}{\bibfnamefont{Y.}~\bibnamefont{Skourski}}, \bibnamefont{and}
  \bibinfo{author}{\bibfnamefont{T.}~\bibnamefont{Hermannsdorfer}}
  (\bibinfo{year}{2009}), \bibinfo{note}{submitted to Crystal Growth and
  Design}.

\bibitem[{\citenamefont{Bewley et~al.}(2006)\citenamefont{Bewley, Eccleston,
  McEwen, Hayden, Dove, Bennington, Treadgold, and Coleman}}]{Merlin}
\bibinfo{author}{\bibfnamefont{R.~I.} \bibnamefont{Bewley}},
  \bibinfo{author}{\bibfnamefont{R.~S.} \bibnamefont{Eccleston}},
  \bibinfo{author}{\bibfnamefont{K.~A.} \bibnamefont{McEwen}},
  \bibinfo{author}{\bibfnamefont{S.~M.} \bibnamefont{Hayden}},
  \bibinfo{author}{\bibfnamefont{M.~T.} \bibnamefont{Dove}},
  \bibinfo{author}{\bibfnamefont{S.~M.} \bibnamefont{Bennington}},
  \bibinfo{author}{\bibfnamefont{J.~R.} \bibnamefont{Treadgold}},
  \bibnamefont{and} \bibinfo{author}{\bibfnamefont{R.~L.~S.}
  \bibnamefont{Coleman}}, \bibinfo{journal}{Physica B}
  \textbf{\bibinfo{volume}{385-86}}, \bibinfo{pages}{1029}
  (\bibinfo{year}{2006}).

\bibitem[{\citenamefont{Perring et~al.}(2009)\citenamefont{Perring, Ewings, and
  Duijn}}]{Horace}
\bibinfo{author}{\bibfnamefont{T.~G.} \bibnamefont{Perring}},
  \bibinfo{author}{\bibfnamefont{R.~A.} \bibnamefont{Ewings}},
  \bibnamefont{and} \bibinfo{author}{\bibfnamefont{J.~V.} \bibnamefont{Duijn}}
  (\bibinfo{year}{2009}), \bibinfo{note}{http://horace.isis.rl.ac.uk, and
  unpublished}.

\bibitem[{\citenamefont{Johnston}(1997)}]{Johnston1997}
\bibinfo{author}{\bibfnamefont{D.~C.} \bibnamefont{Johnston}},
  \emph{\bibinfo{title}{Handbook of Magnetic Materials, Vol 10}}
  (\bibinfo{publisher}{Elsevier Science}, \bibinfo{address}{Netherlands},
  \bibinfo{year}{1997}).

\bibitem[{\citenamefont{Gaft et~al.}(2000)\citenamefont{Gaft, Boulon, Panczer,
  Guyot, Reisfeld, Votyakov, and Bulka}}]{Gaft2000}
\bibinfo{author}{\bibfnamefont{M.}~\bibnamefont{Gaft}},
  \bibinfo{author}{\bibfnamefont{G.}~\bibnamefont{Boulon}},
  \bibinfo{author}{\bibfnamefont{G.}~\bibnamefont{Panczer}},
  \bibinfo{author}{\bibfnamefont{Y.}~\bibnamefont{Guyot}},
  \bibinfo{author}{\bibfnamefont{R.}~\bibnamefont{Reisfeld}},
  \bibinfo{author}{\bibfnamefont{S.}~\bibnamefont{Votyakov}}, \bibnamefont{and}
  \bibinfo{author}{\bibfnamefont{G.}~\bibnamefont{Bulka}}, \bibinfo{journal}{J.
  Lumin.} \textbf{\bibinfo{volume}{87-9}}, \bibinfo{pages}{1118}
  (\bibinfo{year}{2000}).

\bibitem[{\citenamefont{Hohenberg and Brinkman}(1974)}]{Hohenberg1974}
\bibinfo{author}{\bibfnamefont{P.}~\bibnamefont{Hohenberg}} \bibnamefont{and}
  \bibinfo{author}{\bibfnamefont{W.}~\bibnamefont{Brinkman}},
  \bibinfo{journal}{Phys. Rev. B} \textbf{\bibinfo{volume}{10}},
  \bibinfo{pages}{128} (\bibinfo{year}{1974}).

\bibitem[{\citenamefont{Furrer and Gudel}(1977)}]{Furrer1977}
\bibinfo{author}{\bibfnamefont{A.}~\bibnamefont{Furrer}} \bibnamefont{and}
  \bibinfo{author}{\bibfnamefont{H.~U.} \bibnamefont{Gudel}},
  \bibinfo{journal}{Phys. Rev. Lett.} \textbf{\bibinfo{volume}{39}},
  \bibinfo{pages}{657} (\bibinfo{year}{1977}).

\end{thebibliography}

\end{document}